\documentstyle[prl,aps,graphicx,twocolumn]{revtex}

\begin{document}
\wideabs{
\title{Ultra-precise measurement of optical frequency ratios}
\author{J\"orn Stenger, Harald Schnatz, Christian Tamm and Harald R. Telle}
\address{Physikalisch-Technische Bundesanstalt, Bundesallee 100, 38116 Braunschweig, Germany}
\date{\today}
\maketitle
\begin{abstract}
We developed a novel technique for frequency measurement and synthesis, based on the 
operation of a femtosecond comb generator as transfer oscillator. The technique can be used to measure
frequency ratios of any optical signals throughout the visible and near-infrared part of the spectrum.
Relative uncertainties of $10^{-18}$ for averaging times of 100~s are possible.
Using a Nd:YAG laser in combination with a nonlinear crystal we measured the frequency ratio of the 
second harmonic $\nu_{SH}$ at 532~nm to the fundamental $\nu_0$ at 1064~nm,
$\nu_{SH}/\nu_0 = 2.000\,000\,000\,000\,000\,001 \times (1 \pm 7 \times 10^{-19})$. 
\end{abstract}
}

The development of optical frequency comb generators 
based on Kerr-lens mode-locked femtosecond lasers \cite{udem99,didd00} 
has enormously stimulated the field of optical frequency synthesis and metrology.  
Using this technique, the absolute frequencies of a number of narrow transitions in cold atoms or single stored ions 
such as H, Ca, Yb$^+$ or Hg$^+$ have been measured by phase-coherently linking those signals to 
primary cesium-clock controlled hydrogen masers \cite{nier00,sten01,udem01,sten01b}. 
The measurement instabilities approached those of the hydrogen masers, indicating that 
neither the optical frequency standards nor the frequency combs themselves were limiting parts of the setups.

Any absolute frequency measurement is finally limited by the frequency instability of the 
device realizing the unit of frequency, Hertz, such as a radio or microwave reference like the hydrogen maser.
A possibility to avoid this limitation is the measurement of optical frequency {\it ratios}, which
are unitless. 
Thus, frequency ratios for oscillators with better stability than that of the radio or microwave reference 
can be determined with smaller uncertainty than the absolute 
frequencies if a technique is available to realize the frequency ratio without introducing additional noise.

Such a technique is the transfer oscillator concept, which has been realized with a harmonic 
frequency chain \cite{kram99}. However, the measured frequency ratios were restricted to small integer numbers.
In this Letter, we describe a novel technique based on the operation of a femtosecond frequency comb 
generator as a transfer oscillator. Our technique has the capability of generating {\it arbitrary} ratios of any 
optical frequencies throughout the visible and near-infrared part of the spectrum, 
while frequency fluctuations of the comb modes do not enter the measurement but cancel out. 

We demonstrate the superior short-term instability by two measurements: 
first, we measured the frequency ratio of signals from a
single-Yb$^+$-ion frequency standard \cite{tamm00} and from an I$_2$-frequency-stabilized 
Nd:YAG laser \cite{cord00}. 
Second, we used the Nd:YAG laser and measured the frequency ratio of  the second harmonic at 532~nm 
to its fundamental at 1064~nm, thereby testing how accurately the 2:1 frequency ratio is realized
by second harmonic generation.
We demonstrate the capability of our technique of frequency-ratio
measurements with relative uncertainty better than $10^{-18}$.

Kerr-lens mode-locked femtosecond lasers emit a periodic train of short pulses. 
The spectrum of this emission corresponds to a comb of distinct modes, that are
exactly equally spaced due to the strong and fast Kerr-lens mode-coupling mechanism \cite{udem99}. 
Any external signal with frequency $\nu_i$ within the comb
spectrum can be related to a suitable comb mode by detection of the beat-note frequency $\Delta_i$, 

\begin{equation}
\nu_i (t) = \nu_{ceo}(t) + m_i f_{rep}(t) + \Delta_i(t)\,.
\label{eq1}
\end{equation}
 
Here,  $m_i$ denotes the integer order number of the comb mode selected for the beat note, 
$f_{rep}$ the pulse repetition rate, and $\nu_{ceo}$ a so called carrier-envelope offset-frequency that 
accounts for the offset of the entire comb with respect to the frequency zero. 
For absolute frequency measurements all detected frequencies, $\nu_{ceo}$, $f_{rep}$, and $\Delta_i$,
are referred to a microwave-reference frequency $f_{R}$, which may be generated by a primary clock. 

The task of optical frequency metrology in general is the establishment of a phase-coherent link 
between two or more external frequencies $f_{R}, \nu_1, \nu_2, \ldots$, 
while the noise contributions from 
the frequency link, such as the femtosecond comb's parameters $\nu_{ceo}$ and $f_{rep}$, 
are kept as small as possible.
Conventional approaches \cite{jone00} aim to reduce fluctuations of $\nu_{ceo}$ and $f_{rep}$ by stabilization of  
both the group and phase delay of the laser resonator, e.g. with the help of piezo transducers. 
However, as a result of the finite response time of these elements, the servo bandwidth of such servo loops 
is generally not sufficient to reduce the frequency noise of the beat-notes $\Delta_i$ to  
that of the optical signals at $\nu_i$. 
The stability of $f_{rep}$ is particularly demanding, since according to eq.~(\ref{eq1})
this quantity is multiplied by the mode number $m_i$, which is of the order of $10^6$.
As a consequence, the short-term instability of such frequency measurements is limited by 
noise from the microwave reference imposed on the frequency comb.

Our novel approach completely differs in handling the technical frequency contributions. 
We synthesize one radio frequency signal containing all necessary information 
from the external optical signals. This radio frequency is independent of 
$\nu_{ceo}$ and $f_{rep}$ and is thus immune against noise contributions from the femtosecond laser.
The frequency $\nu_{ceo}$ cancels out in real-time signal processing and 
the detection of $f_{rep}$ is not required at all. 

\begin{figure}[ht]
  \centerline{\includegraphics[width=8.2cm]{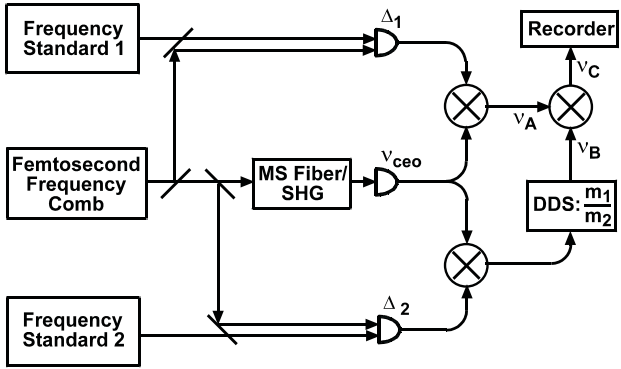}}
  \smallskip
  \caption{Linking two optical frequencies with a femtosecond frequency: real-time signal processing scheme.}
  \label{fig1}
\end{figure}

The setup is sketched in Fig.~\ref{fig1}. 
Three signals are detected: $\nu_{ceo}$ and two beat signals of two external frequency standards with modes
of the femtosecond frequency comb, $\Delta_1$ and $\Delta_2$. 
The sum frequency of $\Delta_1$ and $\nu_{ceo}$ is generated by a mixer.
The sum frequency of $\Delta_2$ and $\nu_{ceo}$ is additionally processed with a 
direct-digital-synthesis IC (DDS). Such a device generates an output signal from an input 
signal with a frequency ratio given by a long digital tuning word. 
It numerically approximates the ratio of the two integers $m_1$ and $m_2$ by $j/2^n$, where $j$ is an integer and 
$n$ the bit-length of the tuning word. 
The resulting possible error is negligible for commercially available $n=48$~bit devices. 
As a result of this signal processing, one obtains 

\[
\nu_A = \nu_{ceo}+\Delta_1 \qquad\mbox{and}\qquad \nu_B = (\nu_{ceo}+\Delta_2) \frac{m_1}{m_2}. \nonumber
\]

Generating the difference frequency of both signals with the help of a mixer, we find using eq.~(\ref{eq1})

\[
\nu_C = \nu_A -\nu_B = \nu_1 -\nu_2  \frac{m_1}{m_2}.
\]

This frequency $\nu_C$ is a measure for the (small) deviation of the optical signal's frequency ratio $\nu_1/\nu_2$
from $m_1/m_2$. Since $\nu_C \ll \nu_1, \nu_2$, the requirements on the radio frequency reference are 
in general not demanding, i.e. it is not necessary to refer $\nu_C$ to a cesium-clock controlled hydrogen maser. 
Alternatively, $\nu_C$ can be referred to one of the optical frequency standards with the help
of the detection of the repetition rate $f_{rep}$. In this case, a self-referenced measurement system is established,
independent of the realization of the unit Hertz. 

The frequency $\nu_C$ carries the full information of the fluctuations of the ratio $\nu_1/\nu_2$, 
but it is independent of the properties of the femtosecond laser, i.e. it is independent
of both the frequency fluctuations of $\nu_{ceo}$ and the repetition rate $f_{rep}$.
The complete cancellation of all femtosecond comb parameters realizes the transfer principle. Essential
requirement is that eq.~(\ref{eq1}) holds for the instantaneous phase angles.
Technically, the cancellation requires a sufficiently fast detection
of the optical signals, but no active stabilization like in the conventional approach mentioned above.
This requirement can be easily fulfilled using fast photodetectos given
the typical Fourier frequency range of the noise of $\nu_{ceo}$ and the beat signals $\Delta_1$ and
$\Delta_2$.

The femtosecond frequency comb is generated 
by our Kerr-lens mode-locked Ti:Sapphire-laser, similar to that described in \cite{sutt99}. 
Pulse duration and repetition rate are $<15$~fs (FWHM) and 100~MHz, respectively. 
Approximately 30~mW of the laser output is coupled into a 10~cm long piece of air-silica microstructure (MS) 
fiber \cite{rank00}, leading to an output spectrum that extends  from about 500~nm to about 1100~nm. 
The carrier-envelope-offset frequency $\nu_{ceo}$ is measured via the beat note of a few ten thousand 
frequency-doubled infrared comb lines at 1080~nm with fundamental lines at 540~nm. 
The beat notes of the external optical signals with comb modes are detected with fast PIN photo diodes.
More technical details related to the setup and the signal processing scheme in Fig.~\ref{fig1} can be found in 
ref.~\cite{tell01}. 

For the two frequency ratio measurements we chose two frequency standards generating three different reference 
frequencies:
\begin{itemize}
\item[i)] the sub-harmonic of the output of a single-ytterbium-ion frequency standard \cite{tamm00} 
        at $\nu_{Yb} = 344\,179\,449$~MHz (871~nm),
\item[ii)] the single-frequency output of a Nd:YAG laser at 
       $\nu_0 = 281\,606\,335$~MHz (1064~nm), that was frequency stabilized to an iodine transition, and 
\item[iii)] the second harmonic of the Nd:YAG laser's emission at $\nu_{SH}= 563\,212\,670$~MHz (532~nm),
       generated by a periodically-poled KTP crystal. 
\end{itemize}

Both the frequency $\nu_{Yb}$ \cite{sten01b} and $\nu_0$ \cite{nevs01} 
were previously measured with respect to a primary cesium atomic clock.

In the first experiment we measured the ratio between $\nu_{Yb}$ and $\nu_0$. 
With respect to Fig.~\ref{fig1}, the beat note of $\nu_{Yb}$ with a comb mode was $\Delta_1$ and that of 
$\nu_0$ was $\Delta_2$. We obtained

\[
\frac{\nu_{Yb}}{\nu_0} = 1.222 \,200 \,739 \,114 \;\times (1 \pm 3 \times 10^{-12}).  
\]

The uncertainty was dominated by the reproducibility of the iodine-stabilized Nd:YAG laser. 

\begin{figure}[ht]
  \centerline{\includegraphics[width=8.2cm]{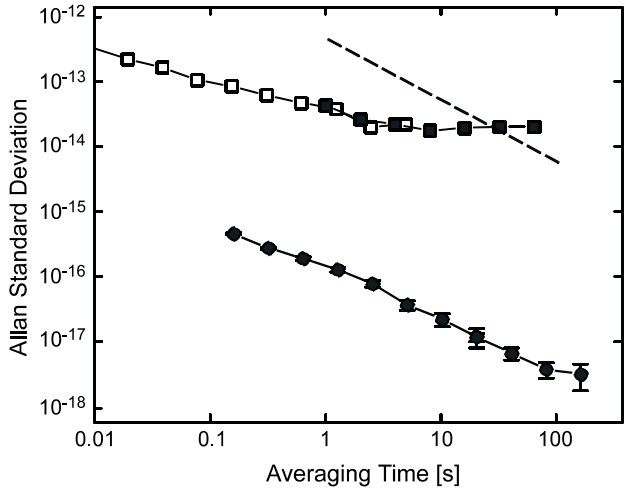}}
  \smallskip
  \caption{Measured Allan standard deviation of two frequency-ratio measurements using the 
Yb$^+$ and Nd:YAG reference frequencies (squares) and the second harmonic and fundamental of a 
Nd:YAG laser (circles). 
The frequency instability of a typical microwave reference (hydrogen maser) is included for comparison 
(dashed line).}
  \label{fig2}
\end{figure}

The Allan standard deviation is shown in Fig.~\ref{fig2} (squares). 
The data for short averaging times (open squares) were derived from the signal 
$\nu_c$ (in Fig.~\ref{fig1}). 
For long averaging times, the real-time processing is not necessarily required.
Instead, the frequency ratio can be calculated offline from averaged values (e.g. 1~s) of
$\nu_{ceo}, \Delta_1$ and $\Delta_2$ (solid squares in Fig.~\ref{fig2}). 
Both data sets agree in the overlapping range between 1 and 8~s. 

Similar uncertainty levels have been obtained by direct comparison of two Nd:YAG lasers, indicating a
limitation of our measurement by the Nd:YAG laser. We concluded that the stability of the reference 
signal pair was not
good enough to investigate any possible noise contribution from the femtosecond comb.
Therefore we used in the second experiment the signal $\nu_0$ and its second harmonic $\nu_{SH}$ 
that was generated in a nonlinear crystal.  
The frequency-doubling process enabled the generation of a pair of reference frequencies 
with completely correlated fluctuations, leading to a fixed frequency ratio 2:1 even for the shortest
averaging times. 
With respect to Fig.~\ref{fig1}, the beat note of $\nu_{SH}$ with a comb mode was $\Delta_1$ and that of 
$\nu_0$ was $\Delta_2$. The DDS was replaced by a frequency doubler.

The experimental result for the frequency ratio of the second harmonic and the fundamental from the Nd:YAG 
laser is 

\[
\frac{\nu_{SH}}{\nu_0} = 2. 000\, 000\, 000\, 000\, 000\, 001 \,\times (1\pm 7\times 10^{-19}).
\]

This measurement sets a new limit on the relative error of the 
1:2 ratio of the frequency doubling process in a nonlinear crystal of smaller than $10^{-18}$. 
This limit is more than five orders of magnitude below previous experimental limits \cite{wyna95}.

Fig.~\ref{fig2} shows the Allan standard deviation for the measurement (circles). 
Indeed, the instability is more than two orders of magnitude smaller than for the $\nu_{Yb}/\nu_0$ measurement,
for which already high quality oscillators were used, and it surpasses even the performance of state-of-the-art
optical oscillators \cite{oate00} substantially. The observed stability can be explained by the estimated fluctuations
of the length difference between the beam paths of the two optical signals. Thus one might expect an even 
better stability when path length fluctuations are technically suppressed. 
 
\begin{figure}[ht]
  \centerline{\includegraphics[width=8.2cm]{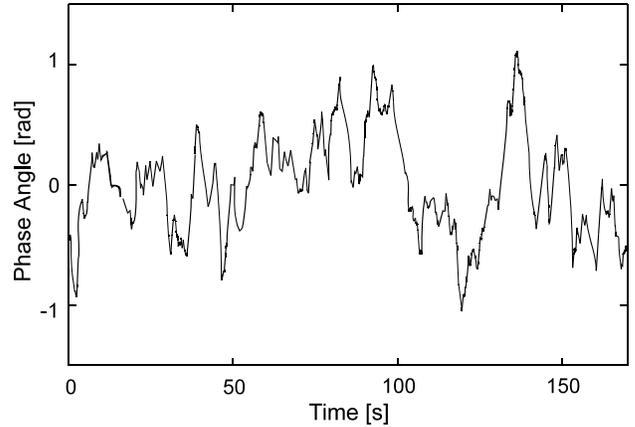}}
  \smallskip
  \caption{Typical relative phase excursions of $\nu_C$ (see Fig.~\ref{fig1}), 
representing the processed second harmonic and fundamental of the Nd:YAG laser. 
In order to avoid base-band detection, the frequencies $\nu_A$ and $\nu_B$  were frequency 
shifted leading to an average value of 25~Hz (arbitrarily chosen) of $\nu_C$.}
  \label{fig3}
\end{figure}

\begin{figure}[th]
  \centerline{\includegraphics[width=8.2cm]{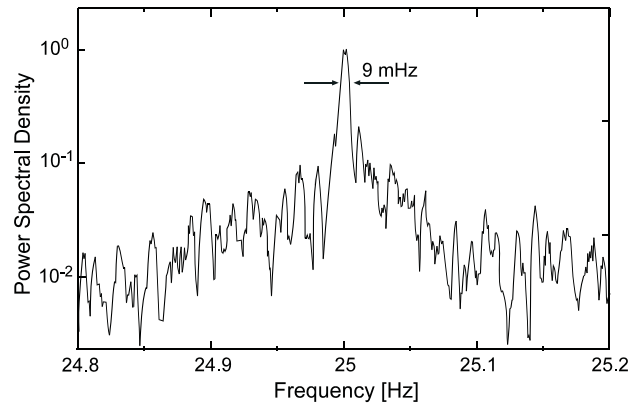}}
  \smallskip
  \caption{Spectrum of $\nu_C$, calculated from the phase excursions in Fig.~\ref{fig3}
by fast Fourier transform. A 900~s data string was used for the calculation.}
  \label{fig4}
\end{figure}

A typical phase record as derived from the signal $\nu_C$ in Fig.~\ref{fig1} is shown in 
Fig.~\ref{fig3}. The phase excursions were smaller than 1~rad for periods of minutes.
From such phase records we obtained by fast Fourier transform a line width of 9~mHz, 
as shown in Fig.~\ref{fig4}. It should be noted that this line was derived from beat notes $\Delta_1$ and $\Delta_2$
with common mode frequency excursions as wide as MHz.

In conclusion, we have demonstrated a novel concept for frequency measurement and synthesis, based on the 
operation of a femtosecond comb generator as transfer oscillator, which does not introduce additional noise. 
It allows to exploit the superior short-term stability of optical frequency standards by the measurement of
frequency ratios. Indeed, the transfer principle using a femtosecond comb is not limited to optical frequencies, 
but can also be applied to radio and microwave frequencies \cite{tell01}.
The capability of such high precision measurements enables new applications in spectroscopy as soon as 
appropriate reference signals from optical transitions can be realized with sufficient frequency stability. 
Interesting questions are effects such as QED-corrections, isotope shifts or relativistic 
corrections to absolute transition frequencies \cite{eide01}.  
Related to the expanding universe, a variation of fundamental constants such as the coupling strengths of 
electromagnetic, weak or strong interaction is expected. 
These constants are indeed accessible with optical spectroscopy \cite{pres95}. 
Relative frequency measurements of optical transitions with different scaling in the finestructure constant 
$\alpha$ are sensitive to temporal changes of the electromagnetic coupling strength. 
Comparison of hyperfine splitting provides an additional sensitivity to changes of the strong interaction strength. 
So far, laboratory experiments have set limits on the variations of about $10^{-14}$~yr$^{-1}$. 
Our frequency-ratio measurement technique, which is capable of linking optical signals with an uncertainty 
of $10^{-18}$ in 100~s, may be part of substantially improved experiments.

We gratefully acknowledge financial support from the Deutsche Forschungsgemeinschaft through SFB~407, 
experimental assistance by N.~Haverkamp and B.~Lipphardt and help for the set--up of the femtosecond laser by 
G.~Steinmeyer and U.~Keller.
We are also indebted to R. Windeler of Lucent Technologies for providing us with the microstructure fiber.


\end{document}